\documentclass[12pt,preprint]{aastex}
\usepackage{epsfig}
\usepackage{natbib}
\usepackage{footnote}
\usepackage{hyperref}
\usepackage[referable]{threeparttablex}

\begin{document}

\title{RX J0513.1$+$0851  AND RX J0539.9$+$0956: TWO YOUNG, RAPIDLY ROTATING, SPECTROSCOPIC BINARY STARS}
\author{Dary Ru\'iz-Rodr\'iguez\altaffilmark{1}, L. Prato\altaffilmark{1}, Guillermo Torres\altaffilmark{2}, L. H. Wasserman\altaffilmark{1}, Ralph Neuh\"auser\altaffilmark{3}}
\altaffiltext{1}{Lowell Observatory, 1400 West Mars Hill
 Road Flagstaff, AZ 86001; dar@lowell.edu}
\altaffiltext{2}{Harvard-Smithsonian Center for Astrophyscis, 60 Garden St.,
Cambridge, MA 02138}
\altaffiltext{3}{Astrophysikalisches Institut und Universit\"ats-Sternwarte, FSU Jena,
Schillerg\"a\ss{}chen 2-3, D-07745 Jena, Germany}

\begin{abstract}

RX J0513.1$+$0851 and RX J0539.9$+$0956 were previously identified as young, low-mass, single-lined spectroscopic binary systems and classified as weak-lined T Tauri stars at visible wavelengths. 
Here we present radial velocities, spectral types, $v \sin i$ values, and flux ratios
for the components in these systems resulting from two-dimensional cross-correlation analysis. These results are based on high-resolution, near-infrared spectroscopy taken with the Keck II telescope to provide a first characterization of these systems as double-lined rather than single-lined. It applies the power of infrared spectroscopy to the detection of cool secondaries; the flux scales as a less steep function of mass in the infrared than in the visible, thus enabling an identification of low-mass secondaries. We found that the RX J0513.1$+$0851 and RX J0539.9$+$0956 primary stars are fast rotators, 60 km s$^{-1}$ and 80 km s$^{-1}$ respectively; this introduces extra difficulty in the detection of the secondary component as a result of the quite broad absorption lines.  To date, these are the highest rotational velocities measured for a pre-main sequence spectroscopic binary. 
The orbital parameters and mass ratios were determined by combining  new visible light spectroscopy with our infrared data for both systems. For RX J0513.1$+$0851, we derived a period of $\sim$4 days and a mass ratio of $q = $ 0.46 $\pm$ 0.01 and for RX J0539.9$+$0956, a period of $\sim$1117 days and a mass ratio
of $q = $ 0.66 $\pm$ 0.01.  
Based on our derived properties for the stellar components, we estimate the luminosities and hence
distances to these binaries at 220 pc and 90 pc. They appear to be significantly closer than previously estimated.

\end{abstract}


\section{Introduction}

Stellar mass is a vital parameter in the life of a star and determines all the evolutionary stages of a single object. Theoretical models for the evolution of young stars are widely used to estimate stellar masses and ages.  For the pre-main sequence (PMS) phase, the theoretical calculations for stars above 1.2 $M_{\odot }$ are approximately consistent with each other. However, for stars less massive than 1.2 $M_{\odot }$, the predicted masses are different for different models \citep{Hillenbrand2004, Simon2008}.   Thus, for low-mass young stars, these estimates are not reliable and few observations exist with which to improve the models.

Close to half of stars with spectral type GKM have a stellar companion \citep[e.g.,][]{Raghavan2010}. In the last two decades, theoretical formation models for binary stars have been evolving. These models assume that the two stars were formed together in the same environment and that their characteristics strongly depend on the initial conditions within the molecular cloud core \citep{Bate2000}. The predictions of these models point out that close binaries are more likely to have similar mass components. However, it is hard to determine the accuracy of these calculations because a direct test requires looking at the statistical properties of a large sample of young double stars, such as the mass ratio distribution.  Existing observations reveal at least some fraction of binary systems with small mass ratios \citep{Prato2002}. However, these results are limited by the small number of young binary systems with accurately measured mass ratios.

The study of individual components in young spectroscopic systems offers further insight into binary formation. The projected rotational velocity of a star provides clues to the angular momentum evolution and may be correlated with the age of the system \citep{Delfosse1998} and the initial conditions of the parent molecular cloud. Some studies of the rotational velocity indicate strong stellar activity \citep{Browning2008, Meibom2007}, as \citet{Alcala1996} suggest with respect to RX J0513.1$+$0851. 

Young spectroscopic binary (SB) systems are fundamental for the understanding of the formation and the evolution of pre-main sequence stars. The dynamical mass ratio can be calculated from the radial velocities (RVs) of a double-lined SB system (SB2).  For systems that are angularly
resolvable or for eclipsing systems, the component stellar masses may be measured with
high accuracy \citep{Schaefer2008, Schaefer2012, Boden2012}.
Many of the known young SB systems are single-lined (SB1). An SB1 does not provide information about the mass ratio; it only allows us to determine the mass function from the calculated orbital parameters, based on the observed primary star RVs. The orbital inclination and masses of the primary and secondary stars are inseparable in the mass function \citep{Mazeh2002}.

Historically, most spectroscopic observations of SBs have been made in visible light. In order to increase the sample size of young SB2s with measured mass ratios, we used the fact that the low-mass secondary component is cooler and redder than the primary component and thus can be detected more efficiently at infrared (IR) wavelengths \citep[e.g.][]{Prato2002A}. Applying this technique, we have observed RX J0513.1$+$0851 and RX J0539.9$+$0956 spectroscopically in the IR.

\citet{Alcala1996} characterized RX J0513.1$+$0851 and RX J0539.9$+$0956 as young, with ages of $\leq $ 10$^{7}$ years, and with spectral types of K3 and K5, respectively. Selected from a survey made by the X$-$ray satellite ROSAT, RX J0513.1$+$0851 showed an emission line with equivalent width of W(H$\alpha$) $=$ $-$19.5 $\pm$ 0.30 \AA, an absorption line of  W(Li) $=$ 0.25 \AA, but an absence of forbidden lines such as [SII] $\lambda$$\lambda$ 4068, 4132, 6717 and 6731 \AA, and [OI] $\lambda$ 6300 and 6363 \AA, characteristic of classical T Tauri stars (CTTs). Based on these results, Alcala et al. classified this system as a weak-lined T Tauri star (WTTs). \citet{Alcala1996} attribute the  H$\alpha$ emission as indicative of a very active chromosphere or flare activity during the spectroscopic observations. In addition, RX J0513.1$+$0851 is not extinguished  and does not present a near-IR excess (Table~\ref{table:Properties}), which indicates a lack of circumstellar or circumbinary dust.

The distance to RX J0539.9$+$0956 was estimated at 460 pc \citep{Alcala1996}, with a temperature of approximately 4,840 K and a stellar luminosity of 15.9 $L_{\odot }$ \citep{Marilli2007}. Mart\'{i}n (1997) found a very short rotation period (Table~\ref{table:Properties}) for this system, possibly symptomatic of an evolved low-mass WTTs approaching the zero age main sequence (ZAMS) with a steady increase of the rotation rate during its contraction. This object was treated as a single star to determine its rotational period \citep{Marilli2007}, a reasonable assumption for visible light observations.  Table~\ref{table:Properties} summarizes the general properties of RX J0539.9$+$0956 \citep{Marilli2007, Alcala1996, Alcala2000}. Near- and mid-IR photometry for RX J0539.9$+$0956, from 2MASS and  Wide-field Infrared Survey Explorer (WISE) data, indicates a lack of significant IR excess and thus little or no circumstellar or circumbinary dust.

In this paper, we present the detection of low-mass secondary components in two rapidly rotating, young spectroscopic binary systems, RX J0513.1$+$0851 and RX J0539.9$+$0956, found by using high-resolution IR spectroscopy to measure the component RVs. These systems are unique within the small sample of about 50 known PMS SB2s because of their very rapid rotation. Section 2 provides a summary of our observations and data reduction. We describe our analysis in section 3 and our results in section 4. Section 5 is a discussion and section 6 lists our conclusions.

\section{Observations and Data Reduction}

\subsection{Visible Light Spectroscopy}

High-resolution visible light spectra for RX J0513.1$+$0851 were taken over a period of 6 years. A total of 32 spectra were obtained between 1996 December and 2002 November with two nearly identical echelle spectrographs \citep{Latham1992} mounted on the 1.5m Tillinghast reflector at the F. L. Whipple Observatory on Mt. Hopkins (AZ) and on the 4.5 m-equivalent Multiple Mirror Telescope (also on Mt. Hopkins) before its conversion to a monolithic mirror. A single echelle order spanning 45 \AA~ was recorded with photon-counting Reticon detectors at a central wavelength of 5187 \AA~ with a resolving power of $R=\lambda /\Delta \lambda \approx 35,000$. The signal-to-noise ratios of these spectra range from 9 to 12, per resolution element of 8.5 km s$^{-1}$. 

Visible light spectra for RX J0539.9$+$0956 were also taken with the two echelle spectrographs on the 1.5m Tillinghast reflector and on the 4.5m-equivalent Multiple Mirror Telescope.  Additional spectra were taken  with the 1.5m Wyeth reflector at the Oak Ridge Observatory (in the town of Harvard, MA). A total of 46 spectra were gathered for this star using the three telescopes between 1992 January and 2009 January; signal-to-noise ratios ranged from 10 to 29 per resolution element and $R=\lambda /\Delta \lambda \approx 35,000$.

\subsection{Near-IR Spectroscopy}

IR spectra of RX J0513.1$+$0851 and RX J0539.9$+$0956 were obtained with the Keck II
telescope and the near-IR spectrometer, NIRSPEC, on Mauna Kea \citep{McLean1998, McLean2000}.
NIRSPEC is a cross-dispersed cryogenic echelle spectrometer using a 1024 x 1024 ALADDIN III InSb array detector. The wavelength range in the near-IR covers 0.95 to 5.4 $\mu$m; the resolving power is 30,000 (8 km s$^{-1}$) using a 0.288 arcsec slit width (2 pixels). The read noise is $\sim$25 electrons rms. The observations were taken in a pattern of four exposures, nodded between two positions on the sky, and hence on the spectrograph slit, in an ABBA sequence for each target in order to subtract the background emission and bad pixels. 

For the detection of the secondary component, the IR spectroscopic technique requires well-defined absorption lines in the stellar spectra. Blending with the telluric lines\footnote{Telluric absorption lines produced  by the absortion of  $O_{3}$, $O_{2}$, and $H_{2}$O are common in IR spectra.} inherent in the Earth's atmosphere can confuse the secondary star signal and also requires observations of a telluric standard star to remove these features.  In order to avoid telluric absorption, the targets were observed in the $H$-band (1.545-1.565 $\mu$m) with a central wavelength $\sim$ 1.555 $\mu$m, corresponding to NIRSPEC order 49. This order provides spectra nearly free of terrestrial absorption \citep{Rousselot2000}.

\begin{figure*}
\begin{center}
\includegraphics[angle=0,width=6.5in]{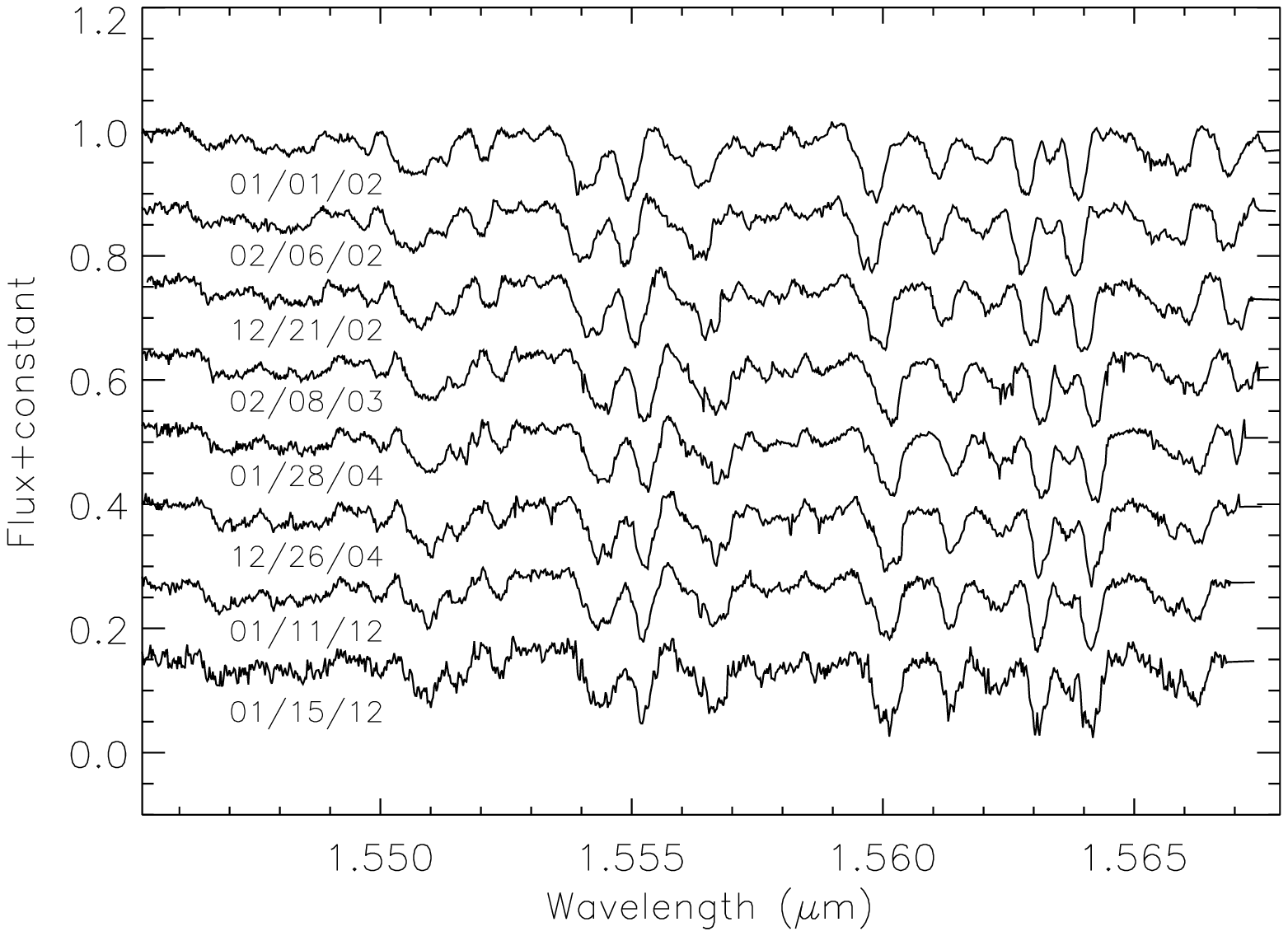}
\caption[RX J0513.1$+$0851 IR Spectra]{Eight epochs of spectra for RX J0513.1$+$0851 in order 49, with barycentric corrections applied; UT dates of the observations are indicated. An arbitrary additive constant was used to offset the spectra for display.}
\label{figure:rxj0513_spectra}
\end{center}
\end{figure*}

\begin{figure*}
\begin{center}
\includegraphics[angle=0,width=6.5in]{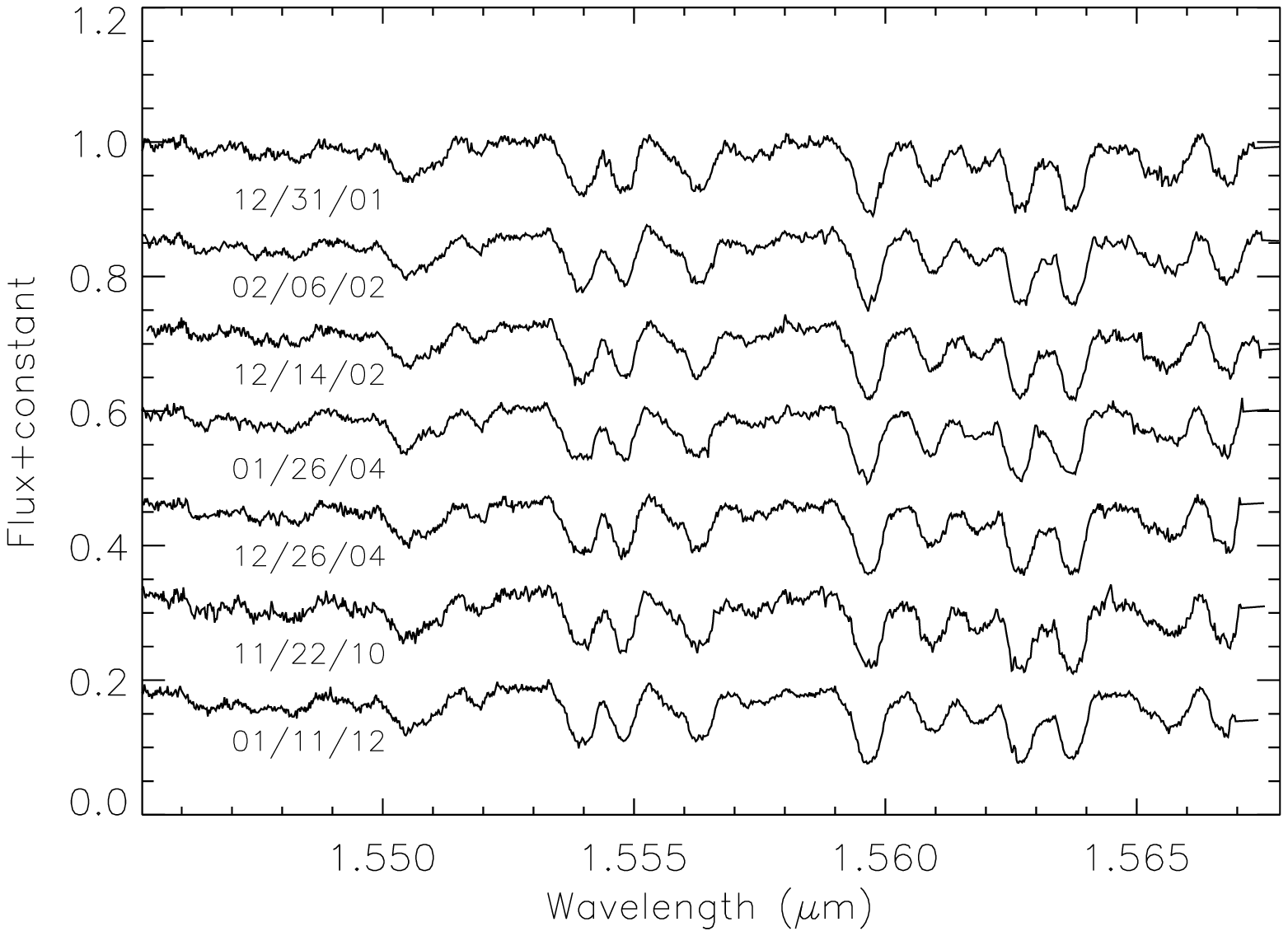}
\caption[RX J0539.9$+$0956 IR Spectra]{Six epochs of spectra for RX J0539.9$+$0956 in order 49, with barycentric corrections applied; UT dates of the observations are indicated.  An arbitrary additive constant was used to offset the spectra for display.}
\label{figure:rxj0539spectra}
\end{center}
\end{figure*}

IR Observations for RX J0513.1$+$0851 were made during a period of 10 years, between 2002 February and 2012 January, and for RX J0539.9$+$0956 between December 2001 and January 2012; specific dates appear in Tables ~\ref{table:RVrxj0513} and  ~\ref{table:RVrxj0539}. All data were reduced using the REDSPEC software package designed at UCLA specially for the analysis of NIRSPEC data. The reduced and barycentric corrected spectra for order 49  are shown in Figures ~\ref{figure:rxj0513_spectra} and ~\ref{figure:rxj0539spectra}.  Further details of the data reduction may be found in \citet{Mace2009}.

\section{Analysis}

\subsection{Visible Light RVs}
Primary star RV measurements based on the visible light observations
of RX J0513.1$+$0851 and RX J0539.9$+$0956 were obtained by
cross-correlation using the IRAF\footnote{IRAF is the Image Reduction and Analysis Facility, a general purpose software system for the reduction and analysis of astronomical data.} task XCSAO \citep{Kurtz1998} and templates from a large library of synthetic spectra
calculated by J. Morse, based on model atmospheres by R. L. Kurucz\footnote{Available at http://cfaku5.harvard.edu.} \citep[see][]{Nordstroem1994, Latham2002}. Of the
four parameters that characterize these templates (effective temperature, rotational
velocity, metallicity, surface gravity), the two that affect the
RVs the most are the temperature and rotational velocity. We
determined these by seeking the best overall match to the observed
spectra as measured by the correlation coefficient averaged over all
exposures of each star \citep[see][]{Torres2002}. For the unknown
metallicity we adopted solar composition, and we also held the surface
gravity fixed at the value $\log g = 4.0$, typical for pre-main
sequence stars. For RX J0513.1$+$0851, we found that a template with an
effective temperature of 5000 K and a projected rotational velocity of
50 km s$^{-1}$ produced the highest cross-correlation coefficient. In the
case of RX J0539.9$+$0956, the best match was found for a temperature
also of 5000 K, and a projected rotational velocity of 60 km s$^{-1}$. Tables
~\ref{table:torres} and ~\ref{table:rvtorres} show the barycentric Julian dates of the observations and the
primary RVs and uncertainties found for RX J0513.1$+$0851 and RX
J0539.9$+$0956, respectively. The uncertainties were originally
determined taking into account both the internal errors as well as the
signal-to-noise ratio of the individual observations. They were then
adjusted through iterations by applying a multiplicative factor until the
reduced $\chi^2$ from a single-lined orbital solution based only on
these velocities was near unity. These are the uncertainties listed in
Tables ~\ref{table:torres} and ~\ref{table:rvtorres}.

\subsection{Infrared Radial Velocities}

To measure the IR RVs (Tables 4 and 5) we used a two-dimensional cross-correlation algorithm
\citep[e.g.,][]{Zucker1994}. The observed spectrum of an SB2 is correlated against a model composed of two template spectra, which represent the primary and the secondary components. This requires high quality template spectra with considerable similarity to the spectra of the stars in the SB2. For this purpose, observed spectra of high signal-to-noise ratio were used as primary and secondary star templates \citep[e.g.,][]{Prato2002}. The template spectra were corrected for barycentric motion and the continua flattened.  It was not possible to determine consistent results on the basis of synthetic template spectra (see Mace et al. 2012).

In addition, to obtain a good match between the template spectra and the observed spectrum, we rotationally broadened the template spectra by convolving them with a line-broadening function \citep{Prato2007,Bender2005,Gray1992,Claret2000}. The secondary to primary light ratios ($\alpha=l_{2}/l_{1}$)  play an important role in maximizing the correlation coefficient  \citep{Zucker1994}. Therefore, the light ratio was determined using the phase at which the component RVs have the greatest difference, and was then held fixed for the analysis of all other phases.

Uncertainties in the IR RVs for both SB systems were initially determined following the same procedure as in \citet{Mace2012}; we found
primary and secondary star RV uncertainties of 4.0 and 5.5 km s$^{-1}$ and 3.0 and 4.0 km s$^{-1}$ for RX J0513.1$+$0851 and RX J0539.9$+$0956, respectively. 
The relatively large uncertainties are likely the result of the high rotational broadening in the absorption lines. 
In the derivation of the double-lined orbital solution, we repeated the same procedure as for the visible light data alone and iteratively multiplied the uncertainties by multiplicative constants until the $\chi^2$ value was close to unity (\S 3.3).

\subsubsection{RX J0513.1$+$0851}

\begin{figure*}
\begin{center}
\includegraphics[angle=0,width=6.5in]{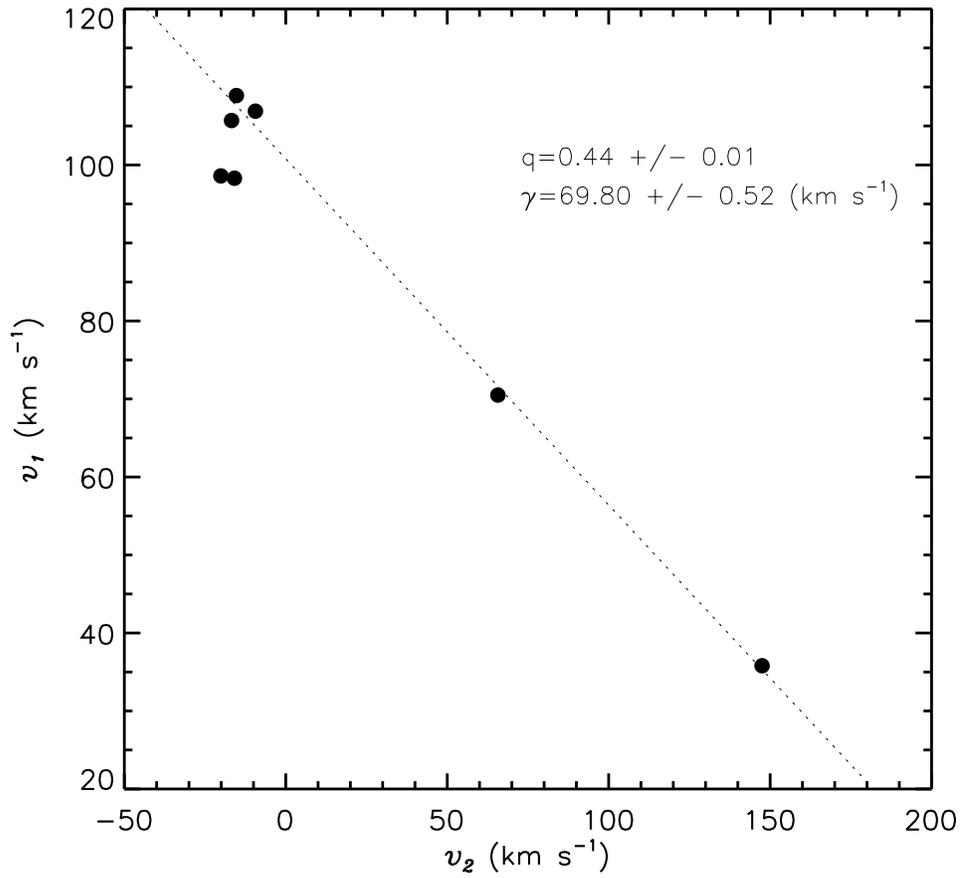}
\caption[RX J0513.1$+$0851 Wilson Plot]{Linear fit to the primary vs. secondary RVs for RX J0513.1$+$0851. Results given for $\gamma$ and $q$ are from the \citet{Wilson1941} analysis.}
\label{figure:RXJ0513_wilson}
\end{center}
\end{figure*}

The observed template spectra \citep{Prato2002} that best matched our RX J0513.05$+$0851 observations were HD 283750 and GL 752\,A, of spectral types K2 and M2, respectively. We found projected rotational velocities, $v \sin i$s, of 60 km s$^{-1}$ for the primary component and
30 km s$^{-1}$ for the secondary component, and an intensity ratio between the two components of 0.20 (Table~\ref{table:derived}). These projected rotational velocities and the intensity ratio maximize the cross correlation coefficient. We estimated an uncertainty of $\pm$ 10 km s$^{-1}$ for the $v \sin i$ values and $\pm$ 0.10 for intensity ratio by varying these parameters until the correlation coefficient decreased measurably.
Table~\ref{table:RVrxj0513} shows the barycentric Julian dates of the observations, along with the RVs and final $\sigma$(RV) values for RX J0513.05+0851. As a preliminary analysis, we made use of the procedure described in \citet{Wilson1941} to derive approximate values of $q = $ 0.44 $\pm$ 0.01 for the mass ratio and $\gamma=68.90 \pm 0.52$ km $s^{-1}$ for the center of mass velocity (Fig.~\ref{figure:RXJ0513_wilson}).

\subsubsection{RX J0539.9$+$0956}

\begin{figure*}
\begin{center}
\includegraphics[angle=0,width=6.5in]{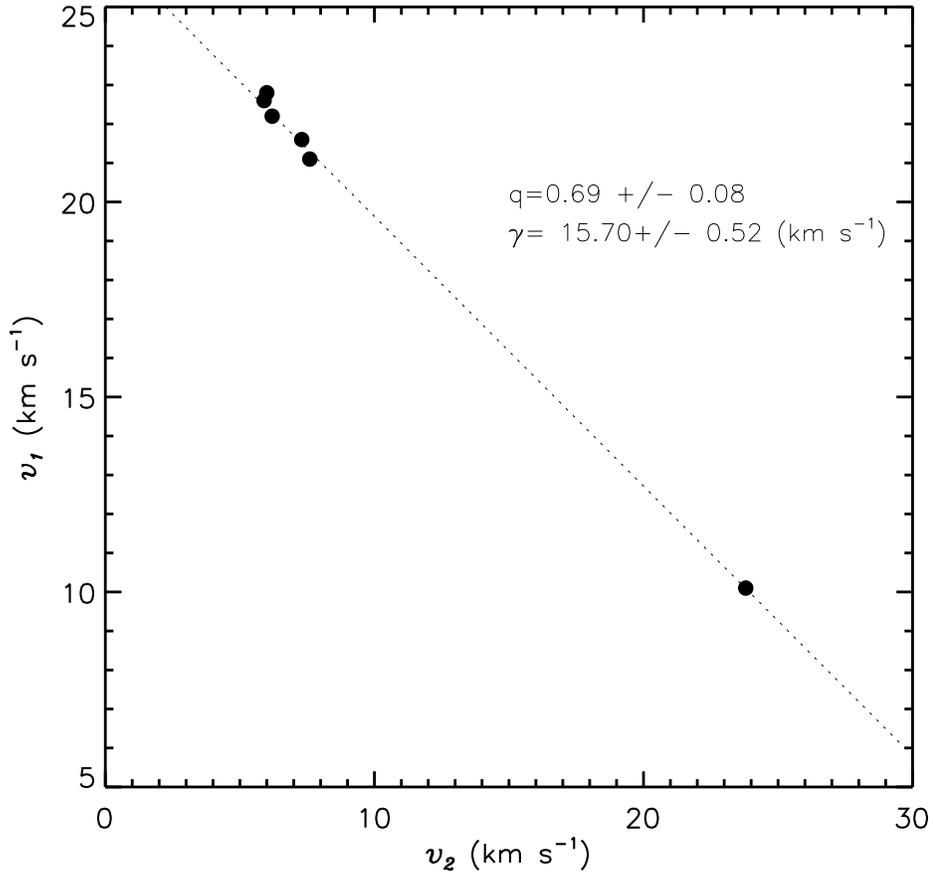}
\caption[RX J0539.9$+$0956 Wilson Plot]{Linear fit of the primary vs. secondary RVs for RX J0539.9$+$0956. Results shown for $\gamma$ and $q$ are from the \citet{Wilson1941} analysis.}
\label{figure:wilsonrxj0539}
\end{center}
\end{figure*}

From Figure~\ref{figure:rxj0539spectra} it is evident that the absorption lines of
RX J0539.9$+$0956 are very broad, meaning that there is a high rotational velocity in the dominant  stellar component. Cross-correlation of the spectra for each epoch
yielded a K5 (GL 1094) with
$v \sin i=80$ km s$^{-1}$ and an M3 (GL 15\,A) with $v \sin i=40$ km s$^{-1}$ for the primary and secondary, respectively. The best fit was found with an M3/K5 intensity ratio of 0.25 (Table~\ref{table:derived}). Uncertainties of $\pm$ 10 km s$^{-1}$ and $\pm$ 0.10 for the $v \sin i$ values and intensity ratio, respectively, were found as for RX J0513.1$+$0851. Table~\ref{table:RVrxj0539} shows the IR RVs, final $\sigma$(RV) values, and barycentric dates of the observations for RX J0539.9$+$0956. A preliminary analysis was performed following  \citet{Wilson1941}; we plotted the six epochs of the primary versus secondary RVs (Figure ~\ref{figure:wilsonrxj0539}) and found a mass ratio of $q=$ 0.69 $\pm$ 0.08 and a center-of-mass velocity $\gamma=$15.7  $\pm$ 0.9 km s$^{-1}$. 

We note a similar inconsistency in the magnitude of the $v \sin i$ values between the visible light and IR observations as in \citet{Mace2012}; the IR values are systematically larger. This difference possibly results from how $v \sin i$ is taken into account in the suite of cross-correlation templates. For the visible light templates, synthetic spectra were created with rotation inherent in the model atmosphere. To simulate a range of $v \sin i$ values in the observed IR templates, a rotation kernel was convolved with the template spectra \citep[e.g.,][]{Prato2007}.  Although it is unclear why this might result in discrepant values for the $v \sin i$ measured in the $V$- and $H$-bands, it seems unlikely that the difference has a physical origin.

\subsection{Results}

We used the Levenberg-Marquardt approach \citep{Press1992} to provide a least-squares orbital fit on the basis of the measured RVs. We excluded IR spectrum taken for RX J0513.1$+$0851 on UT 2002 Jan 1 from our analysis because, in spite of the high signal-to-noise (Figure 1), the secondary star RV was completely inconsistent with our solution based on all other epochs.  In order to adjust the measured visible light and IR RVs so that the reference frame zero points from the two data sets were consistent, we found orbital solutions on the basis of the combined data with an extra free parameter, $\Delta \gamma$, which indicates the difference in the center of mass RV between the visible and IR RVs. We then corrected all the RVs to a common center of mass velocity by adding this constant, $\Delta \gamma$, to all the measured IR RVs in order to reference the results to the frame of the visible light data. For  RX J0513.1$+$0851 this difference in $\gamma$ was $-$0.5 km s$^{-1}$ and for RX J0539.9$+$0956 it was $-$0.96 km s$^{-1}$ (Table~\ref{table:op12}). Combining the visible light and corrected IR RVs produced the best orbital parameters. Table~\ref{table:op12} lists the orbital period ($P$, in days), eccentricity ($e$), time of periastron passage ($T_0$), longitude of periastron ($\omega$, in degrees), velocity
semi-amplitude for the primary and secondary components ($K_1$, $K_2$, in km s$^{-1}$), and center of mass velocity ($\gamma$, in km s$^{-1}$). Derived quantities are the projected semi-major axis of the primary and the secondary ($a_{1,2}$ sin $i$), the mass ratio ($q=M_{2}/M_{1}$), and the minimum masses ($M_{1}$ sin$^3 i$ and $M_{2}$ sin$^3 i$). We cannot determine $M_{1}$, $M_{2}$ or $i$ independently. 

As noted in $\S$ 3.1 and  $\S$ 3.2, the best orbital solutions were obtained with a series of iterations, multiplying first the visible light RV uncertainties, then the visible light plus IR primary star RV uncertainties, and finally the IR secondary star RV uncertainties by constants such that the
visible-only single-lined solution, the visible$+$IR single-lined solution, and finally the
full visible+IR double-lined solution $\chi^2$ per degree of freedom was approximately unity.  Respectively,
the constants found for RX J0513.1$+$085 were 1.25, 0.98, and 0.9.  For  RX J0539.9$+$0956 
we found 1.2, 0.5, and 0.77.  The final, multiplied uncertainties used in the non-linear fit are shown in Tables ~\ref{table:torres}, ~\ref{table:rvtorres}, ~\ref{table:RVrxj0513} and ~\ref{table:RVrxj0539}. Uncertainties in the orbital parameters were determined from the orbital solution found using the Levenberg-Marquardt approach.

\begin{figure*}
\begin{center}
\includegraphics[angle=0,width=6.5in]{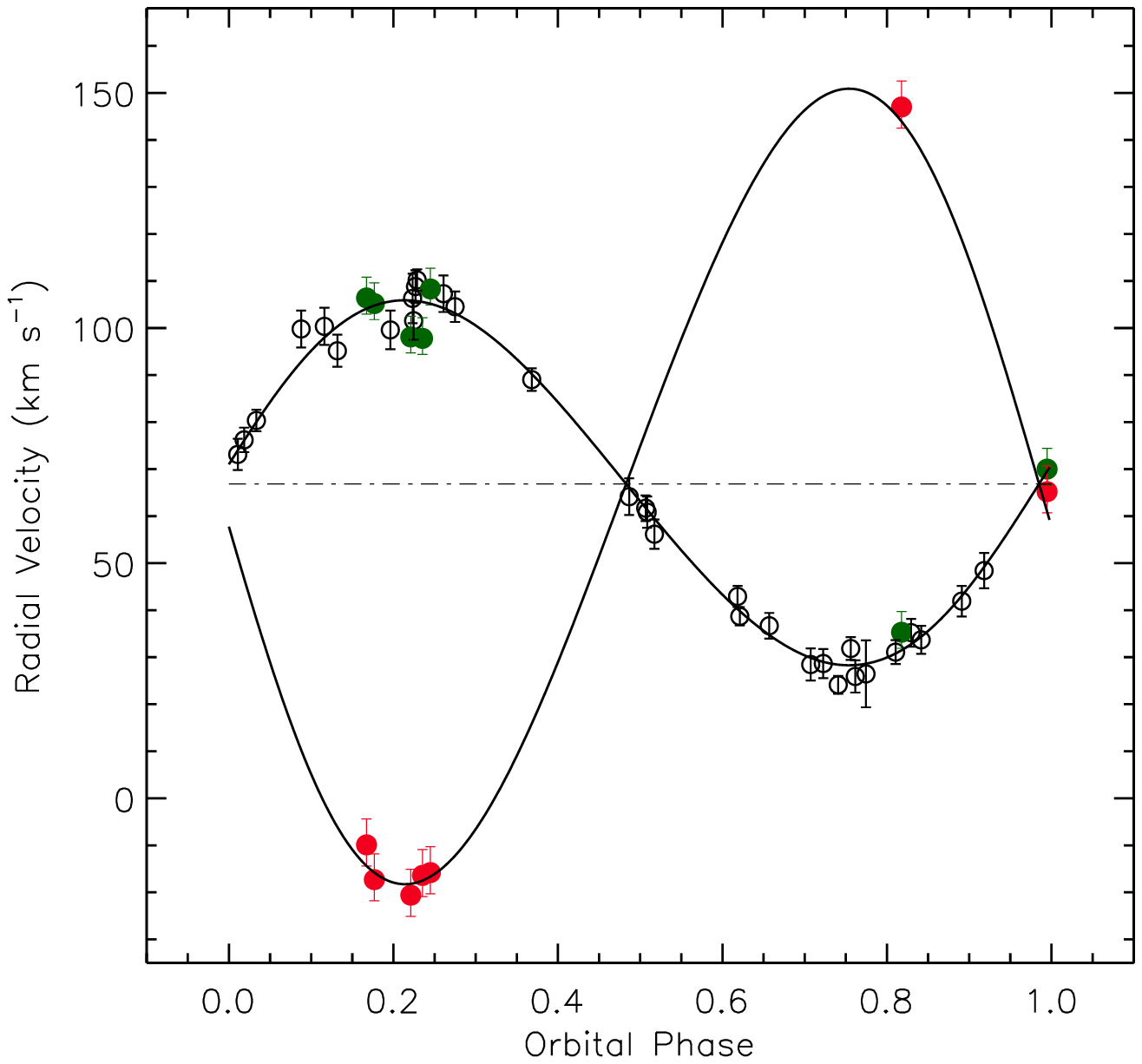}
\caption[RX J0513.1$+$0851 Orbital Phase]{Radial velocity as a function of phase for RX J0513.1$+$0851. The black open circles represent the primary star visible light data, the green circles the primary star IR data, and the red circles represent the secondary star IR data. The best orbital fit is shown as a solid line for the primary and secondary components.The dashed horizontal  line represents the system's center-of-mass velocity. Uncertainties in the primary IR RVs are 3.9 km s$^{-1}$ and in the secondary star RVs are 5 km s$^{-1}$. }
\label{figure:RXJ0513_phase}
\end{center}
\end{figure*}

The mass ratio and the center of mass velocity for RX J0513.1$+$085 given in Table~\ref{table:op12} are consistent with those derived from the \citet{Wilson1941} approach to within 1-2$\sigma$. The RV versus phase curve is plotted in Figure ~\ref{figure:RXJ0513_phase}, showing the primary star velocities in the visible and IR, and secondary star velocities in the IR, as a function of phase.  The observed RVs are almost all within 1$\sigma$ of the calculated orbit.

\begin{figure*}
\begin{center}
\includegraphics[angle=0,width=6.5in]{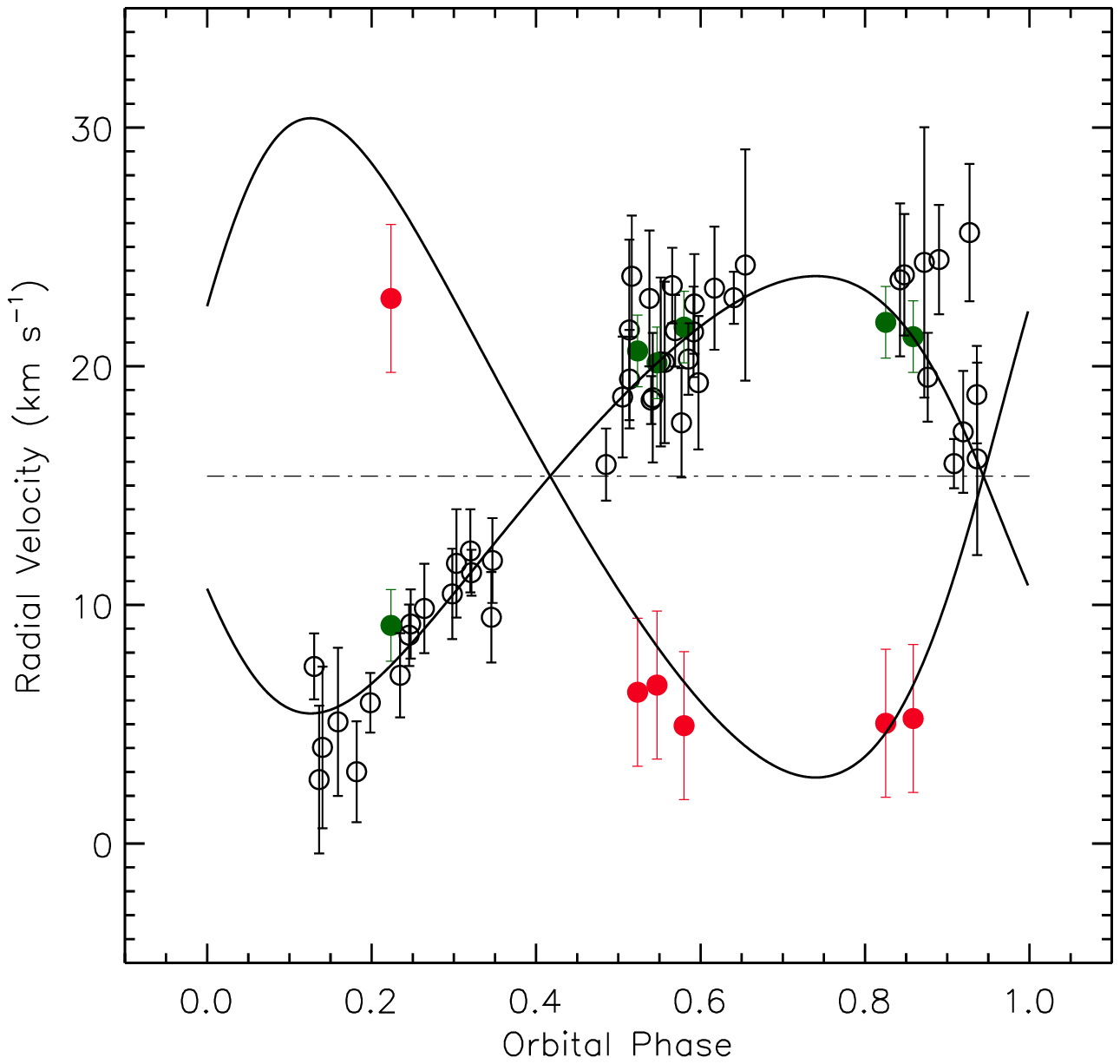}
\caption[RX J0539.9$+$0956 Obital Phase]{Radial velocity as a funcion of phase for RX J0539.9 +0956. The black open circles represent the primary star visible data, the green circles the primary star IR data and the red  circles represent the secondary star IR data. The best orbital fit is shown as a solid line for the primary and secondary components. The dashed horizontal  line represents the system's center-of-mass velocity. Uncertainties in the primary RVs are 1.5 km s$^{-1}$ and the secondary star RVs are 3.1 km s$^{-1}$.}
\label{figure:orbitalrxj0539}
\end{center}
\end{figure*}

Similarly, for RX J0539.9$+$0956, the mass ratio and the center-of-mass velocity derived from the \citet{Wilson1941} approximation are consistent with those in Table~\ref{table:op12} to within 1-2$\sigma$. The orbital solution is plotted in Fig.~\ref{figure:orbitalrxj0539}; the observed RVs are typically within 1$\sigma$ of the orbital solution except for a few of the visible light RVs.

\section{Discussion}

\subsection{RX J0513.05+0851}

\begin{figure*}
\begin{center}
\includegraphics[angle=0,width=6.0in]{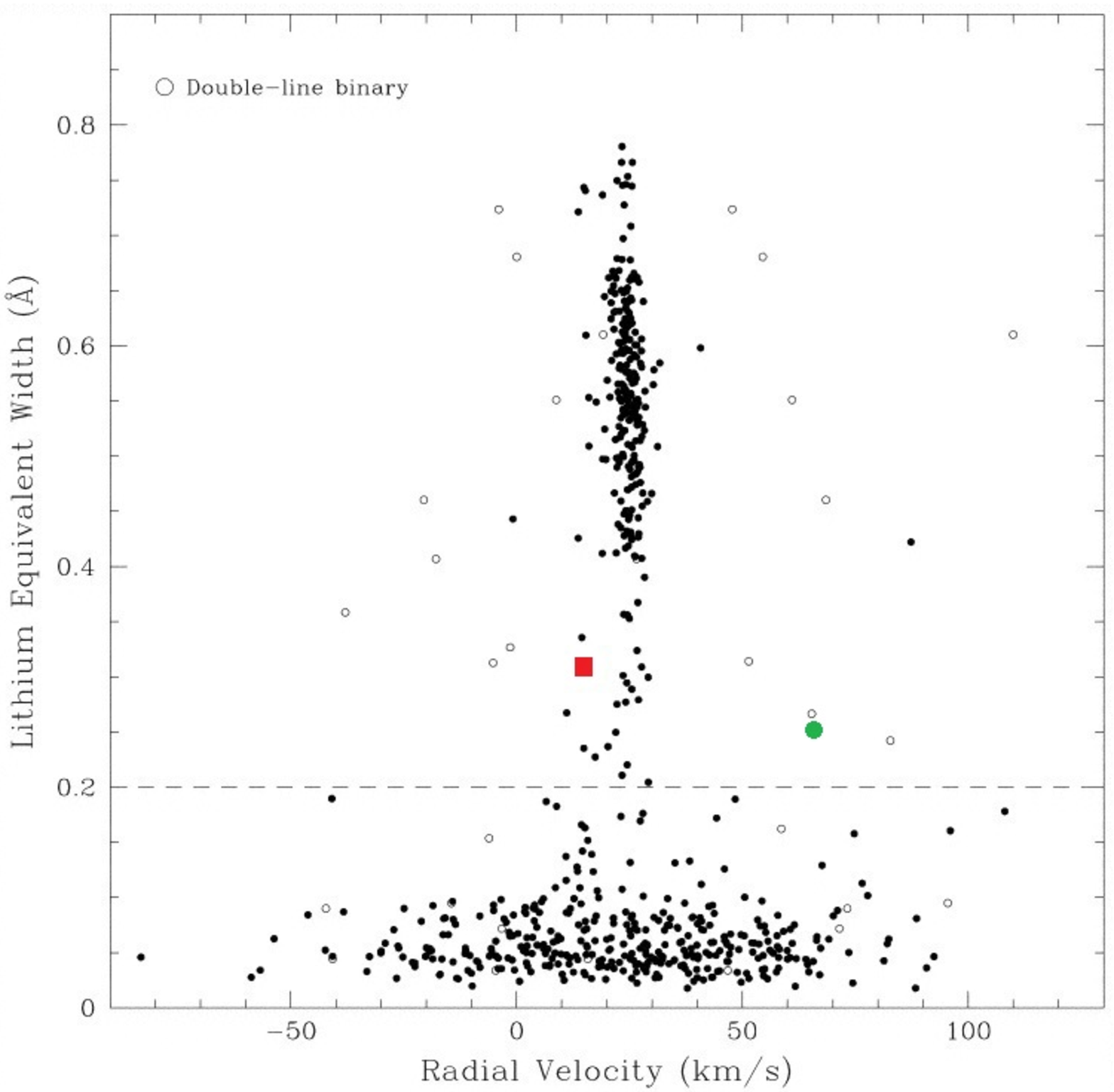}
\caption[$\lambda$ Ori Association]{Figure 4 from \citet{Dolan2001} showing the discrimination between field stars and low-mass members of the $\lambda$ Ori association. The dashed line indicates an empirical cutoff of $W_{\lambda }(Li)\geq$ 0.2 \AA, identifying young stars with strong lithium absorption. The small circles are from \citet{Dolan2001}; open symbols are binaries. The green circle and red square represent the center of mass RVs for RX J0513.05+0851 and  RX J0539.9$+$0956, respectively.  These symbol sizes are at least 10 times larger than the uncertainties of these points (see Table ~\ref{table:op12}).}
\label{figure:dolan}
\end{center}
\end{figure*}

\begin{figure*}
\begin{center}
\includegraphics[angle=0,width=6.5in]{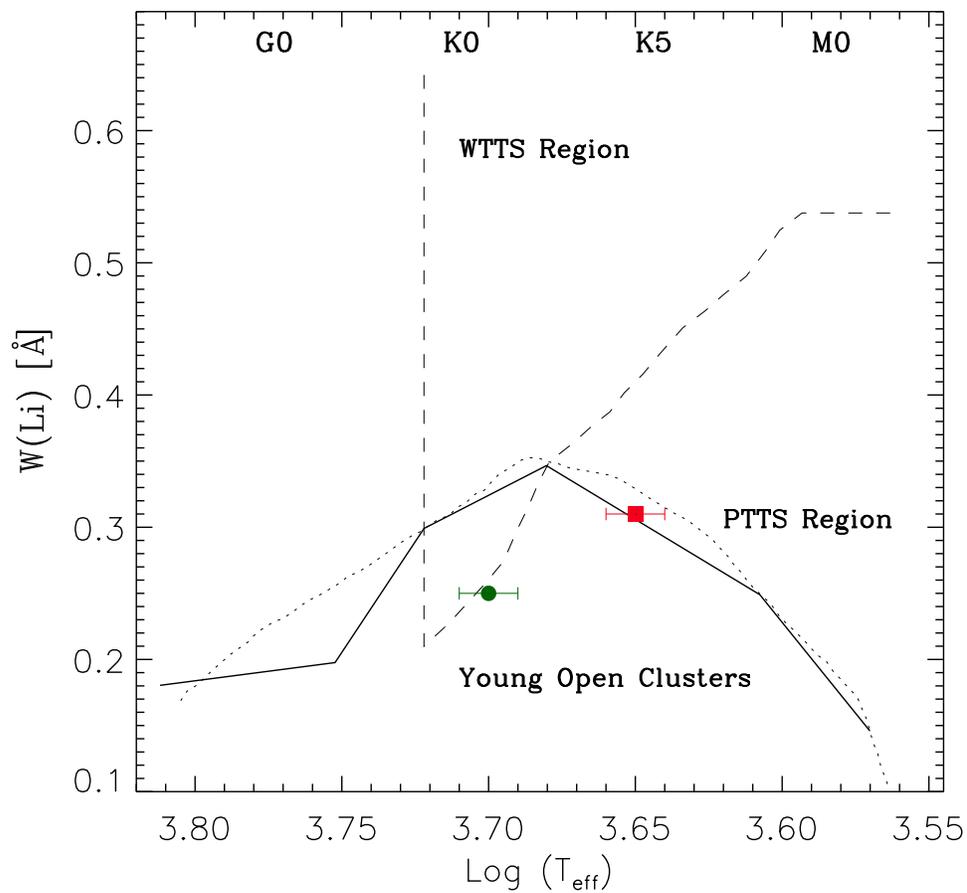}
\caption[Lithium Equivalent Width vs Effective Temperature]{Lithium equivalent width versus effective temperature for single stars. The thick and dotted lines represent the upper envelope for young open clusters \citep{Martinandmagazzu1999}. The dashed line indicates the WTTs and PTTs regions as described in \citet{Martin1997}. The green circle and red square represent RX J0513.05 +0851 and  RX J0539.9$+$0956, respectively.}
\label{figure:alcala2000}
\end{center}
\end{figure*}

\begin{figure*}
\begin{center}
\includegraphics[angle=0,width=6.5in]{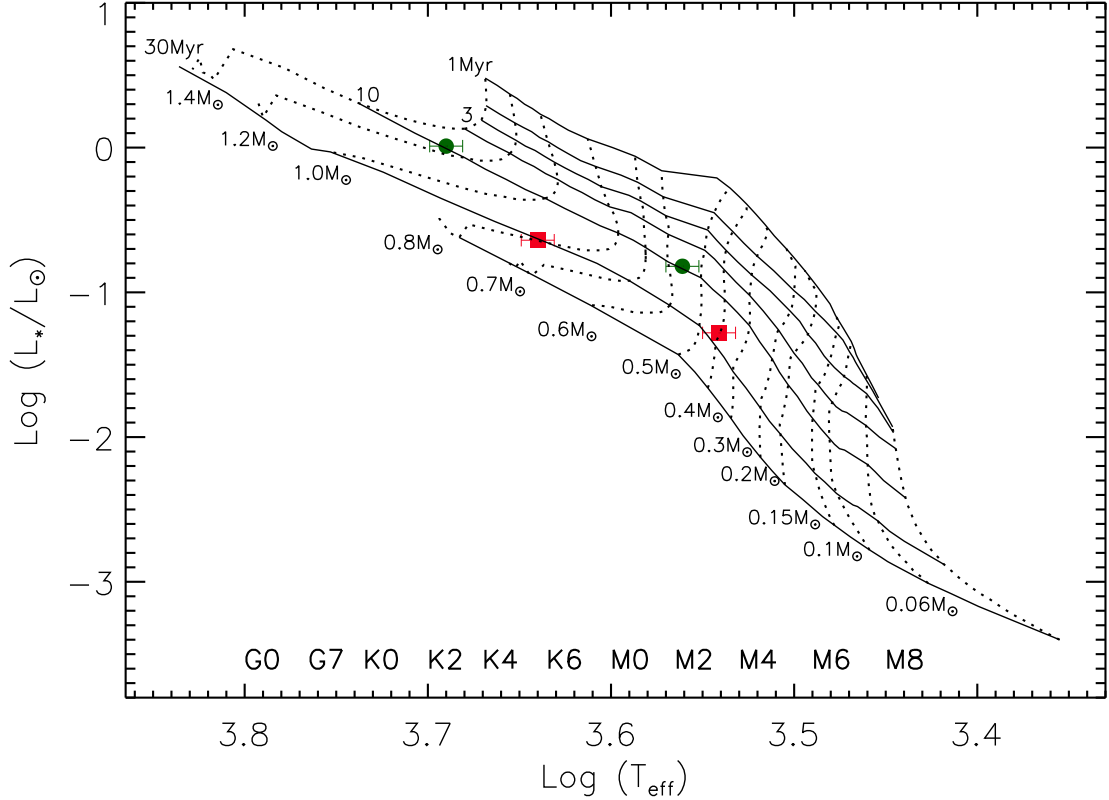}
\caption[Baraffe Theoretical Tracks]{ A Hertzsprung-Russell diagram with the locations of the RX J0513.05+0851 and RX J0539.9$+$0956 components (\S 4) compared to the solar metallicity theoretical evolutionary models of \citet{Baraffe1998}. $T_{\rm eff}$s and luminosities for each component are discussed in the text. The evolutionary tracks are shown as dashed lines with mass in units of M$_{\odot}$. For M $\leqslant 0.6 M_{\odot }$ the adopted mixing length is $\alpha _{mix}$= 1 corresponding to a helium mass fraction  of Y$=$ 0.275, and for M $>$ 0.6 M$_{\odot }$ the mixing length is $\alpha _{mix}$ = 1.9 and Y $=$ 0.282.  Solid lines are isochrones representing ages of 1, 2, 3, 5, 10, 30, and 100 Myr  (from top to bottom). The green circles and red squares represent primary and secondary stars of RX J0513.05+0851 and  RX J0539.9$+$0956, respectively.}
\label{figure:luhman}
\end{center}
\end{figure*}

We measured the primary and secondary stars in the RX J0513.1$+$0851 SB by cross-correlating a K2 and an M2 with respective projected rotational velocities ($v \sin i$) of $\sim$ 60 km s$^{-1}$ and $\sim$ 30 km s$^{-1}$. This factor of two difference in their projected velocities could result from differences in accretion or in magnetic braking as a function of mass \citep{Konopacky2012, Gomez2009}.  It is also possible that the stellar rotation axes are not aligned, although this
would be surprising in such a short-period ($\sim$4 day) system.
\citet{Alcala1996} located this system at $\sim$ 400 pc, corresponding to the distance of the Orion cloud complex \citep{Jeffries2007}. Using 2MASS JHK magnitudes to determine an $A_{v}$ of $\sim$ 0.38, as in \citet{Prato2003}, and assuming that RX J0513.1$+$0851 is in Orion, the total luminosity is  $L\sim$ 5.5 $L_{\odot}$. This is very large for these spectral types, even if
we suppose the system is only a few Myr old. In addition, RX J0513.05$+$0851 has a high center of mass velocity, $\gamma$ $\sim$ 66.8 $\pm$ 0.6 km s$^{-1}$, which is not characteristic of Orion or of the known young, nearby stellar associations and clusters \citep{zuckerman2004}. These discrepancies lead us  the following considerations:

\begin{itemize}

\item RX J0513.1$+$0851 is probably not part of the Orion cloud complex. \citet{Dolan2001} associated strong lithium sources identified in $\lambda$ Ori with RVs of the systems and found a strong peak at $\sim$ 24.5 km s$^{-1}$ with a dispersion of 2.3 km s$^{-1}$ (Figure ~\ref{figure:dolan}).  Previously, the RV range found for some Orion member stars was 21 to 30 km s$^{-1}$  \citep{Hartmann1986}. With a center of mass velocity of 66.8 $\pm$ 0.6 km s$^{-1}$, RX J0513.1$+$0851 is many sigma outside this range.  RX J0513.1$+$0851 is likely closer than 400 pc and part of a different population.  We examined the proper motion of RX J0513.1$+$0851 \citep{Hog2000}, and compared it to $\sim$a dozen putative 32 Ori members \citep{Mace2009, Shvonski2010, Mamajek2012}. Although RX J0513.1$+$0851 is in the same part of the sky as 32 Ori, its center of mass velocity and proper motion are distinct; thus, we conclude that it is not a member of this group either.

\item  For RX J0513.1$+$0851, W(Li) $=$ 0.25 \AA  ~\citep{Alcala1996}, on the low end of the range of lithium-rich stars in Orion with values between 0.2 and 0.8 \AA  ~\citep{Dolan2001, Alcala2000}. Figure ~\ref{figure:alcala2000} shows lithium equivalent width versus effective temperature for single stars. Using the derived spectral type for the primary star of K2 and assuming the system's W(Li) measurement is associated with the primary star, the location of RX J0513.1$+$0851 primary in Figure~\ref{figure:alcala2000} is consistent with the upper envelope for young open clusters adopted by \citet{Martinandmagazzu1999}. Traditionally, low-mass PMS objects are found in the vicinity of star-forming regions. However, there is evidence for young Li-rich, magnetically active, late-type WTTs which are widely dispersed \citep{Neuh1997}. If RX J0513.1$+$0851 is a PMS system and part of an open cluster, it is likely still moving out from the region where it was formed. Alternatively, RX J0513.1$+$0851 could have formed locally in small cloud-let, which has dispersed \citep{Feigelson1996}.

\item Authors such as \citet{Sterzik1995} and \citet{Reipurth2003} have suggested a scenario of possible ejections of low-mass stars from the molecular parent cloud. This is based on the dynamical interactions that occur within small groupings, specifically the scenario of three stars which pass close  enough that their motions are strongly perturbed relative to one and other. The interchange of energy within the three body system will cause one of the stars to acquire more kinetic energy, resulting in the ejection of the lightest member or least stable member. The total energy is conserved and the kinetic energy of the escaping star is released by the formation of a binary star system. However, the ejected star with the higher kinetic energy should be single, thus this does not satisfactorily explain the high center of mass velocity for the RX J0513.1$+$0851 binary. 

\item In order to estimate a distance we rely on the models of \citet{Baraffe1998}, shown in Figure~\ref{figure:luhman}. Although these and other tracks are not totally accurate, and our project aims to provide fundamental data to improve these models, we use them here as a tool to estimate some of the stellar properties.  The reasonableness and consistency of our results may also be indicative of the accuracy of the models\footnote{The \textit{GAIA} mission will return astrometric data which will enable a precise measurement of distance to these systems within the next 3$-$4 years.}.
Given the spectral types determined for RX J0513.1$+$0851 in our cross-correlation analysis, K2 and M2, we adopt the intermediate temperature scale of \citet{Luhman1999} and estimate the corresponding effective temperatures ($T_{\rm eff}$). Uncertainties for $T_{\rm eff}$ are $\sim$145 K, equal to $\pm$1 spectral subclass. Projecting the values of log($T_{\rm eff}$) in a vertical line (not plotted) across the \citet{Baraffe1998} tracks in Figure~\ref{figure:luhman} and requiring that the two binary components be coeval, we found that the 10 Myr isochrone is most consistent with the intensity ratio, 0.2, determined
from the two-dimensional cross-correlation, given the values of the component $T_{\rm eff}$. The combined $T_{\rm eff}$ values and the 10 Myr isochrone yield a mass of 1.2 $\pm$ 0.1 $M_{\odot }$ for the primary star and 0.58 $\pm$ 0.1 $M_{\odot }$ for the secondary star.  Fortuitously, these values result in a mass ratio almost identical to our value of 0.46. The adopted $T_{\rm eff}$ along with the 10 Myr
isochrone lead to estimated luminosities for the components of $L_1 =$ 1.12 $L_{\sun}$ and $L_2 =$ 0.16 $L_{\sun}$, and a total luminosity for the system
of $L_{\rm tot} =$ 1.28 $L_{\sun}$. Inverting the process described in \citet{Prato2003} for finding the luminosity based on the near-IR magnitudes, A$_v$, K-band excess, and distance, we now use 2MASS magnitudes, A$_v=$ 0.38, and zero K-band excess to estimate a distance to RX J0513.1$+$0851 of 220pc.

\end{itemize}

\subsection{RX J0539.9$+$0956} 

Results from our analysis revealed spectral types of K5 and M3 and
rapid rotation for both the primary component of RX J0539.9$+$0956,  $v \sin i$ $\sim$ 80 km s$^{-1}$, and for the secondary, $v \sin i$ $\sim$ 40 km s$^{-1}$. The detection of the secondary component was quite challenging because of the high component rotational velocities. The difference between the rotational velocities, a factor of two, could be attributed to the initial formation conditions or to uneven early accretion \citep{Konopacky2012, Gomez2009}, as for RX J0513.1$+$0851. It is also possible, given the $\sim$ 3 year period of this system, that the stellar inclinations are distinct.

\citet{Marilli2007} found a luminosity for RX J0539.9$+$0956 of $L=$ 15.9 $L_{\odot}$ assuming this system is a member of Orion and a mean distance to the Orion complex of 460 pc; the relatively large lithium equivalent width
(Table~\ref{table:Properties}) is consistent with the corresponding assumption of youth. However, stars of spectral types, K5 and M3, are not massive enough to emit such a high luminosity, even if we assume an age of 1 Myr. We thus take into account similar considerations and follow a parallel procedure as describe in $\S$ 4.1 for RX J0513$+$0851. We draw the following conclusions:

\begin{itemize}

\item The center of mass RV of RX J0539.9$+$0956 is 15.40 $\pm$ 0.33 km s$^{-1}$, only moderately inconsistent with the RV of the
$\lambda$ Ori association in Orion of 24.5 km s$^{-1}$ with dispersion of 2.3 km s$^{-1}$ \citep{Dolan2001} (Figure~\ref{figure:dolan}) and the RV range from 21 to 30 km s$^{-1}$ found for some Orion member stars by \citet{Hartmann1986}. Thus we cannot necessarily exclude this system from membership in the Orion region given the relatively large scatter in the association members' RVs.  A stronger argument against membership in Orion is that its $\sim$400 pc distance would necessitate an unphysically large luminosity, given the spectral types of the stars in RX J0539.9$+$0956.  As in the case of RX J0513.1$+$0851, we examined the proper motion of RX J0539.9$+$0956 \citep{Hog2000} compared to those of 32 Ori members \citep{Mace2009, Shvonski2010, Mamajek2012}. In spite of the spatial coincidence with 32 Ori, the RX J0539.9$+$0956 center of mass velocity and proper motion are distinct thus it is unlikely to be a 32 Ori member.

\item Even if it is not located in Orion, the presence of a small lithium equivalent width in RX J0539.9$+$0956 indicates youth.  Figure ~\ref{figure:alcala2000} suggests that RX J0539.9$+$0956 is a post T Tauri star (PTTs), as described by \citet{Martin1997}. If the transition from WTTSs to PTTSs takes place at around 10 Myr of age \citep{Martin1998}, a PTTs could be defined as a low-mass young star that has not yet reached the ZAMS. This points to a relatively young age for RX J0539.9$+$0956. 
\citet{Briceno1997} argued that the presence of a detectable lithium equivalent width is still consistent with the zero age main sequence (ZAMS). They claimed that the majority of the RASS lithium-rich stars found in the general direction of star forming regions are not PMS. Another scenario for the presence of lithium was proposed by \citet{Alcala1996} who posit that for late-type stars it is possible to overestimate the W(Li) in low-resolution spectra with blending between nearby lines, especially for fast rotators. A third possibility is that stellar rotation may affect lithium depletion, hence the preservation of lithium content at high $v \sin i$ rates \citep{Balachandran2011}. 

\item To estimate the RX J0539.9$+$0956 luminosity and distance, we again relied on the models of
\citet{Baraffe1998}.  From \citet{Luhman1999} we found values of $T_{\rm eff}$ of 4365 K and 3468 K
for the spectral types determined by cross-correlation, K5 and M3.  The isochrone most
consistent with these $T_{\rm eff}$s, coeval binary components, and our observed flux ratio, 0.25, has an age of about 30 Myr (Figure~\ref{figure:luhman}). The theoretical models predict for the primary star a mass of 0.8 $\pm$ 0.1 $M_{\odot }$  and 0.4 $\pm$ 0.1 $M_{\odot }$ for the secondary star. The corresponding mass ratio, 0.5, is within $\sim$1$\sigma$ of our dynamical results. Following the same analysis as for RX J0513.1$+$0851, we find that the corresponding component and thus total
luminosities ($L_{1} =$ 0.234 $L_{\odot }$, $L_{2} = $ 0.07 $L_{\odot }$, and $L_{total} =$ 0.3 $L_{\odot }$) imply a distance of 90 pc. 

\end{itemize}

\section{Summary}

We used high-resolution IR spectroscopy in order to measure the Doppler shift in the absorption lines of both components in two young SB2s, RX J0513.1$+$0851 and RX J0539.9$+$0956. On the basis of high-resolution, visible light spectroscopy these systems were initially identified by \citet{Alcala1996} as SBs and were thought to be located in Orion cloud complex
(d$\approx$400pc). Using two-dimensional cross-correlation, we measured the IR RVs for the primary and the secondary of both binaries, combined these values with the earlier epochs of visible light data, and found a double-lined orbital solution for each system. The best spectral types found in the cross-correlation of the IR data are a K2 with $v \sin i$ = 60 km s$^{-1}$ as the primary and an M2 with $v \sin i$ = 30 \ km s$^{-1}$ as the secondary for RX J0513.1$+$0851, and a K5 with $v \sin i$ = 80 km s$^{-1}$ as the primary and an M3 with $v \sin i$ = 40 km s$^{-1}$ as the secondary for RX J0539.9$+$0956. To date, these are the most rapidly rotating young SBs known.  From the orbital solution, we found a mass ratio of $q =$ 0.46 $\pm$ 0.02 and period of $\sim$4 days for RX J0513.1$+$0851 and $q = $ 0.66 $ \pm$ 0.11 and a period of $\sim$1117 days for RX J0539.9$+$0956. 
On basis of the \citet{Baraffe1998} pre-main sequence models and our derived stellar properties, we estimated the luminosities, $L= $ 1.28 $L_{\odot }$ and
$L=$ 0.3 $L_{\odot }$, and distances, $\sim$ 220 and 90 pc, for RX J0513.1$+$0851 and RX J0539.9$+$0956, respectively. These estimates are important because to date we do not have an accurate method to measure the distance and space motion of these targets.
These luminosities are significantly lower than would be expected if RX J0513.1$+$0851 and RX J0539.9$+$0956 were located at a distance of $\ga$400 pc, but are consistent with the low masses and relatively young ages of the component stars. 
RX J0513.1$+$0851 and RX J0539.9$+$0956 are significant within the sample of known young SB2s because of their high rotational velocities. This provides us with fundamental clues about the possible initial conditions in the formation of a binary system and the subsequent evolutionary phases.\\

\begin{acknowledgments}
We thank the Keck staff for their support of this science, in particular T. Stickel, J. Rivera, S. Magee, G. Puniwai, 
C. Wilburn, G. Hill, and B. Schaefer. We are also grateful to P. Berlind, M. Calkins, J. Caruso, R. Davis,
D. Latham, A. Milone, R. Stefanik, and J. Zajac for their assistance
in obtaining the spectroscopic observations with the CfA facilities,
and to R. Davis for maintaining the database of echelle observations. We thank R. Mathieu for permission to reproduce Figure 7.
We thank the referee for detailed comments which have improved the presentation of this manuscript.
This research was funded in part by a NASA Keck PI Data Award, administered
by the NASA Exoplanet Science Institute, and NSF Grants AST-0444017 and AST-1009136 (to LP). 
GT acknowledges partial support from the NSF through grant AST-1007992. Some of the data described herein were taken
on the Keck II telescope with time granted by NOAO, through the Telescope
System Instrumentation Program (TSIP). TSIP is funded by NSF. 
Some data presented herein were obtained at the W.M. Keck
Observatory from telescope time allocated to the National Aeronautics
and Space Administration through the agency's scientific partnership
with the California Institute of Technology and the University of
California. The Observatory was made possible by the generous
financial support of the W.M. Keck Foundation.
This work made use of the SIMBAD reference database, the NASA
Astrophysics Data System, and the data products from the Two Micron All
Sky Survey, which is a joint project of the University of Massachusetts
and the Infrared Processing and Analysis Center/California Institute
of Technology, funded by the National Aeronautics and Space
Administration and the National Science Foundation. 
We recognize and acknowledge the
significant cultural role that the summit of Mauna Kea
plays within the indigenous Hawaiian community and are
grateful for the opportunity to conduct observations
from this special mountain.
\end{acknowledgments}

\clearpage
\bibliographystyle{apj}
\bibliography{Template}

%%%%%%%%%%%%%%%%%%TABLES%%%%%%%%%%%%%%%%%%%%%%%%%%

\begin{deluxetable}{lccc}
 \tablewidth{0pt}
   \centering
 \tablecaption{General Properties of RX J0513.1$+$0851 and  RX J0539.9$+$0956}
    \tablehead{\colhead{Property} & \colhead{RX J0513$+$0851} & \colhead{RX J0539$+$0956} 
}
    \startdata
R.A.\tablenotemark{a} (J2000)  & 05:13:05.819   &   05:39:56.5  \\
Decl.\tablenotemark{a} (J2000)  & +08:51:31.44 &+09:56:37 \\
$B$\tablenotemark{b}  (mag)&13.80 &11.62 $\pm$ 0.1 \\
$V$\tablenotemark{b} (mag)&12.59 $\pm$ 0.01&11.17 $\pm$ 0.1 \\
$R$\tablenotemark{b} (mag)&11.88 $\pm$ 0.01&--- \\
$J$\tablenotemark{a}  (mag)&10.16 $\pm$ 0.02&9.04 $\pm$ 0.03\\
$H$\tablenotemark{a}  (mag)&9.51 $\pm$ 0.03&8.52 $\pm$ 0.04 \\
$K$\tablenotemark{a} (mag)&9.35 $\pm$ 0.02&8.37 $\pm$ 0.02 \\
W1\tablenotemark{c} (mag)&9.251 $\pm$ 0.022&8.267 $\pm$ 0.023 \\
W2\tablenotemark{c} (mag)&9.218 $\pm$ 0.020&8.271 $\pm$ 0.020 \\
W3\tablenotemark{c} (mag)&9.177 $\pm$ 0.034&8.273 $\pm$ 0.030 \\
W4\tablenotemark{c} (mag)&9.111 $\pm$ $-$&8.367 $\pm$ $-$ \\
P$_{rot}$\tablenotemark{d} (days)&---& 0.764 \\
W(Li)\tablenotemark{e}  (\AA) &0.250 $\pm$ 0.001&0.310 $\pm$ 0.001 \\
W(H$\alpha$)\tablenotemark{e} (\AA)&$-$19.5 $\pm$ 0.1&$-$0.19 $\pm$ 0.10 \\
\enddata

\tablenotetext{a}{2MASS All-Sky Point Source Catalog.}
\tablenotetext{b}{SIMBAD Astronomical Database.}
\tablenotetext{c}{WISE All-Sky Data Release.}
\tablenotetext{d}{\citet{Martin1997}.}
\tablenotetext{e}{\citet{Alcala1996}.}
\label{table:Properties}

%\tablerefs{(1) \citet{Cutri2003}, (a)\citet{Hog2000} , (2) \citet{Schmidt2009} , (3) \citet{Martin1997}, (4) %\citet{Alcala1996}}
\end{deluxetable}

\begin{deluxetable}{c c c c c }
    \tablewidth{0pt}
   \centering
  \tablecaption{Visible Light RVs for RX J0513.1$+$0851 and Residual from the Orbital Solution}
    \tablehead{\colhead{BJD} & \colhead{v$_{1}$} & \colhead{$\sigma$} & ($O - C$)& Orbital Phase \\
    \colhead{} & \colhead{(km s$^{-1}$)} & \colhead{(km s$^{-1}$)} & \colhead{(km s$^{-1}$})}
    \startdata
2450446.8298  & 28.44  & 3.41 & $-$1.28  & 0.7073 \\
2450499.8055  & 41.93  & 3.25 &$-$0.97  & 0.9909 \\  
2450770.9482  & 89.03  & 2.41& $-$1.28 &  0.3682 \\
2450800.8537  & 31.08  & 2.55& 0.46 &  0.8105 \\  
2450820.6648  & 24.12  & 1.90 & $-$4.14 &  0.7407\\ 
2450821.7499  & 73.12  &3.33& $-$0.91  & 0.0108 \\  
2450822.7546  & 107.3  & 3.89 & 2.91  & 0.2608\\  
2450885.6701  & 48.42  & 3.75 &$-$0.67 &  0.9181 \\   
2450916.6229  & 38.69  & 1.94 &$-$1.02 &  0.6211 \\  
2450920.6303  & 42.92  & 2.24 & 2.78  & 0.6184 \\  
2451088.9495  & 61.69  & 2.73 &$-$0.03 &  0.5067 \\ 
2451089.9511  & 31.84  & 2.41 &3.71 &  0.7559 \\  
2451093.0105  & 56.16  &3.11 &$-$3.29  & 0.5173 \\ 
2451121.0152  & 64.13  & 3.89 &$-$1.90  & 0.4866 \\   
2451186.7341  & 33.69  & 2.96 & $-$0.49 &  0.8416\\
2451268.6481  &108.82 & 3.49& 2.88 &  0.2269 \\ 
2451569.6377  & 95.16  & 3.41 &$-$5.34  & 0.1318 \\  
2451621.6985 &  99.80  & 3.92 & 6.85  & 0.0878\\ 
2451832.9361 &  36.69  & 2.73& 2.08 &  0.6569 \\   
2451947.7242  &106.30 & 5.26 & 0.29 &  0.2233\\   
2452180.9923 & 104.53 & 3.25& 1.27 &  0.2749\\  
2452212.9363  &101.55 &4.05 &$-$4.44 &  0.2246\\  
2452214.9375 &  28.66  & 3.10&$-$0.20 &  0.7226 \\ 
2452240.9503  & 99.60  & 4.10 & $-$6.23 &  0.1962\\ 
2452335.6423  & 25.91  & 3.41 &$-$2.25 &  0.7615 \\ 
2452365.6482   &110.22  & 2.26  & 4.32 &  0.2288\\
2452537.9814& 100.36& 3.94& 2.22 & 0.1161 \\
2452568.9758  & 35.22 & 2.95 & 2.62 &  0.8294 \\
2452593.9054 & 80.34& 2.28 & 0.22 &  0.0334 \\
2452595.8144 & 60.85 & 3.30 & $-$0.48 &  0.5085 \\
2452596.8832 & 26.45 & 7.11 & $-$1.99 &  0.7745 \\
2452601.8819& 76.21 & 2.6 & 0.08  & 0.0185
\enddata
\label{table:torres}
\end{deluxetable}

\begin{deluxetable}{ccccc}
    \tablewidth{0pt}
   \centering
  \tablecaption{Visible Light RVs for RX J0539.9$+$0956 and Residuals from the Orbital Solution}
    \tablehead{\colhead{BJD} & \colhead{v$_{1}$} & \colhead{$\sigma$} & ($O - C$)& Orbital Phase\\
    \colhead{} & \colhead{(km s$^{-1}$)} & \colhead{(km s$^{-1}$)}& \colhead{(km s$^{-1}$)}}
    \startdata
    2448644.6376  & 10.46 & 1.90 & 0.05 &  0.2982  \\
    2448670.7602   &11.35 & 0.96  &$-$0.05 &  0.3216  \\
    2448697.7073  &  9.48  & 1.90  &$-$2.94  & 0.3457 \\
    2448914.9247   &18.58  &1.01 & $-$1.33  & 0.5401  \\
    2449767.7615  & 11.74  & 2.27 & 1.12  & 0.3033 \\
    2449786.7094  & 12.26  & 1.74 &0.92 &  0.3203  \\
    2449816.6208  & 11.86  & 1.78  & $-$0.62 &  0.3470 \\
    2449970.9944  & 15.88  & 1.51  &$-$2.14 &  0.4852 \\
    2450002.9480  & 19.46  &2.06 & 0.43 &  0.5138 \\
    2450005.9717  & 23.77   & 2.54 & 4.64  & 0.5165 \\
    2450029.9794  & 22.84  &2.84 & 3.00  & 0.5380  \\
    2450060.9001 &  23.38  & 1.58 &2.67  & 0.5657  \\
    2450064.9193  & 21.49  & 1.50 & 0.68  & 0.5692 \\
    2450082.8519  & 20.30   & 1.50 & $-$0.97 &  0.5853 \\
    2450089.8773  & 21.44   &1.90 & $-$0.01 &  0.5916  \\
    2450090.9414   & 22.61 & 2.09 & 1.14 &  0.5925 \\
    2450144.6798 &  22.87   & 1.09& 0.24  & 0.6406 \\
    2450408.0042  & 19.54   & 1.86 &$-$0.73  & 0.8763 \\
    2450443.8363  & 15.92   & 1.03& $-$2.27 &  0.9084 \\
    2450474.8193 &  18.81  & 2.04 & 2.76 &   0.9361  \\
    2450768.0190 &   5.90   & 1.25 & $-$0.76  & 0.1985 \\
    2450820.8430 &   8.73   &1.28 & 0.45 &  0.2458  \\
    2450822.8680  &  9.20    &1.45 & 0.85 &  0.2476 \\
    2451110.8597  & 18.71  &2.53 & $-$0.03 &  0.5053 \\
    2451151.8781 &  18.68  & 2.71  & $-$1.29 &  0.5420 \\
    2451235.7212 &  23.27  & 2.58  & 1.17 &  0.6171 \\
    2451277.5270  & 24.24  &4.85 & 1.34  & 0.6545  \\
    2451487.8112  & 23.62  & 3.20 & 1.72 &  0.8427  \\
    2451520.7699 &  24.35  & 5.66 & 3.85  & 0.8722 \\
    2451540.7491  & 24.47  & 2.29  & 5.03 &  0.8900 \\
    2451573.5729  & 17.25  & 2.56 & $-$0.12 &  0.9194  \\
    2451592.6500 &  16.12 & 4.03  & 0.11 &  0.9365 \\
    2451815.8649  &  2.68  & 3.10 & $-$2.80 &  0.1362 \\
    2451866.8464  &  3.01  & 2.11  & $-$3.20 &  0.1819\\
    2452699.5888  & 25.60  & 2.88 & 8.83 &  0.9271  \\
    2452937.8303  &  4.03   & 3.38  & $-$1.48 &  0.1403\\
    2452958.8358 &   5.10   & 3.11 & $-$0.64  & 0.1591 \\
    2453354.7803 &  21.52  & 3.78 & 2.50 &  0.5135 \\
    2453402.6936 &  20.16  & 3.38 & $-$0.26 &  0.5564  \\
    2453425.6159  & 17.63  & 2.29  & $-$3.41 &  0.5769 \\
    2453448.5899  & 19.31  & 2.80  & $-$2.29  &  0.5974 \\
    2454043.9197  &  7.42   &1.38  & 1.96 &  0.1302\\
    2454160.7274   & 7.05   & 1.76 &$-$0.82 &   0.2348 \\
    2454193.6679  &  9.85   &1.87 & 0.84 &  0.2642 \\
    2454514.8194 &  20.18  &3.54 & $-$0.10 &  0.5517 \\
    2454845.8486 &  23.83   &2.54 & 2.14  & 0.8479 \\
 \enddata
\label{table:rvtorres} 
\end{deluxetable}

\begin{deluxetable}{cccccc}
    \tablewidth{0pt}
   \centering
  \tablecaption{IR RVs for RX J0513.1$+$0851}

    \tablehead{ \colhead{BJD}& \colhead{v$_{1}$\tablenotemark{a}} & \colhead{$(O-C)$} &\colhead{v$_{2}$\tablenotemark{b}} & \colhead{$(O-C)$} &Orbital Phase \\
    \colhead{} &\colhead{(km s$^{-1}$)} &\colhead{(km s$^{-1}$)} & \colhead{(km s$^{-1}$)} & \colhead{(km s$^{-1}$)}}
    \startdata
2452311.75855 &35.8 & 4.00  & 147.5& 3.04 &0.8177  \\
2452629.91432& 70.5& 0.43 & 65.7 & 4.29 &  0.9947 \\
 2452678.86556&105.7& 0.25 & -16.8 &-1.43 &  0.1768 \\
 2453032.74867& 108.9&2.99 & -15.3 &0.84 &  0.2450\\
 2453365.95440&106.9&2.10 & -9.4 &4.54 &  0.1673   \\
2455937.93153& 98.3& -7.90 & -15.9  &1.09 &  0.2354  \\
 2455941.89231&98.6& -7.93& -20.1 &-2.40 &  0.2211 \\
 \enddata
\label{table:RVrxj0513} 
\tablenotetext{a}{Primary star RV uncertainty is 3.9 km s$^{-1}$.}
\tablenotetext{b}{Secondary star RV uncertainty is 5.0 km s$^{-1}$.}
\end{deluxetable}

\begin{deluxetable}{cccccc}
    \tablewidth{0pt}
   \centering
  \tablecaption{IR RVs for RX J0539.9$+$0956}

    \tablehead{ \colhead{BJD}& \colhead{v$_{1}$\tablenotemark{a}} & \colhead{$(O-C)$} &\colhead{v$_{2}$\tablenotemark{b}} & \colhead{$(O-C)$} &Orbital Phase \\
    \colhead{} &\colhead{(km s$^{-1}$)} &\colhead{(km s$^{-1}$)} & \colhead{(km s$^{-1}$)} & \colhead{(km s$^{-1}$)}}
    \startdata
 2452274.92156 &21.1 &0.01& 7.6 &-1.62 &  0.5471 \\
 2452311.77812&22.6&0.51 & 5.9 &-1.82  & 0.5801  \\
2452622.99356& 22.2&0.04 & 6.2& -1.41 &  0.8586 \\
 2453030.94592& 10.1&1.67 & 23.8&-4.50  & 0.2237 \\
2453365.99928& 21.6&1.28 & 7.3  &-3.08 &  0.5235   \\
2455937.92445& 22.8&-0.69& 6.0 &0.39  & 0.8253   \\
 \enddata
\label{table:RVrxj0539} 
\tablenotetext{a}{Primary star RV uncertainty is 1.5 km s$^{-1}$.}
\tablenotetext{b}{Secondary star RV uncertainty is 3.1 km s$^{-1}$.}
\end{deluxetable}

\clearpage
\begin{deluxetable}{ccccc}
 \tablewidth{0pt}
   \centering
 \tablecaption{Cross-Correlation Derived Properties}
    \tablehead{\colhead{Object} & \colhead{Spec. Type} & \colhead{$T_{\rm eff}$\tablenotemark{a}}& \colhead{ $v \sin i$}& \colhead{Flux Ratio} \\
 \colhead{} & \colhead{} &\colhead{(K)} & \colhead{(km s$^{-1}$)} }

    \startdata
RX J0513.1$+$0851&  K2\tablenotemark{b} & 5010 $\pm$ 145&60 $\pm$ 10 & 0.20 $\pm$ 0.10\\& M2\tablenotemark{c} & 3630 $\pm$ 145 & 30 $\pm$ 5\\
\hline\\
RX J0539.9$+$0956&  K5\tablenotemark{b} & 4365 $\pm$ 145& 80 $\pm$ 10& 0.25 $\pm$ 0.10\\
& M3\tablenotemark{c}  & 3468 $\pm$ 145 &  40 $\pm$ 5 \\
\enddata

\label{table:derived}
\tablenotetext{a}{Spectral Type -- $T_{\rm eff}$ conversion based on \citet{Luhman1999}. }
\tablenotetext{b}{Primary Component.}
\tablenotetext{c}{Secondary Component.}
\end{deluxetable}

\begin{deluxetable}{ccc}
    \tablewidth{0pt}
   \centering
  \tablecaption{Spectroscopic Orbital Solutions from Combined Visible Light and IR Data for RX J0513.1$+$0851 and  RX J0539.9$+$0956}
     \tablehead{\colhead{Orbital parameter} & \colhead{RX J0513.1$+$0851}& \colhead{RX J0539.9$+$0956}} 
    \startdata
P (days) & 4.01829 $\pm$ 0.00006&1117.39 $\pm$ 5.34\\
$\emph{$T_{0}$}$ (BJD) & 2452525.46 $\pm$    0.16& 2451663.6 $\pm$ 57.0 \\
$\emph{e}$ & 0.06 $\pm$  0.02 &0.20 $\pm$ 0.05  \\
$K_{1}$ (km s$^{-1}$)& 39.0 $\pm$    0.7&9.2 $\pm$    0.7\\
$K_{2}$ (km s$^{-1}$)& 84.6 $\pm$   2.7&13.8 $\pm$    2.1\\
${\omega}$ (deg) & 275.8 $\pm$   14.5& 115 $\pm$ 18 \\
$\gamma$ (km s$^{-1}$)& 66.8 $\pm$ 0.6& 15.40 $\pm$    0.33\\
$\Delta \gamma$\tablenotemark{a} (km s$^{-1}$) &$-$0.5 $\pm$ 1.5&$-$0.96 $\pm$ 0.71\\
$a_1$ sin $i$ (10$^{6}$ km) & 2.15  $\pm$ 0.04  &138.0 $\pm$ 9.6 \\
$a_2$ sin $i$ (10$^{6}$ km) &4.66  $\pm$  0.15& 208.0  $\pm$  31.6 \\
$q = M_2/M_1$ & 0.46 $\pm$ 0.02&0.66 $\pm$ 0.11\\
$M_{1}$ sin$^3 i$  (M$_{\odot}$)& 0.53 $\pm$ 0.04&0.79 $\pm$ 0.27\\
$M_{2}$ sin$^3 i$  (M$_{\odot}$)& 0.25 $\pm$ 0.01&0.53 $\pm$ 0.12\\
%$\sigma_{v}$ (km s$^{-1}$)&3.4&2.2\\
%$\sigma_{p}$ (km s$^{-1}$)&5&0.9\\
%$\sigma_{s}$ (km s$^{-1}$)&2.3&2.5\\
\enddata
\tablenotetext{a}{Factor to adjust visible and IR RVs to the same reference frame (\S 3.3).}
\label{table:op12}

 \end{deluxetable}

%%%%%%%%%%%%%%%%%%%%%%%%%%%%%FIGURES%%%%%%%%%%%%%%%%%%%%

\clearpage

\clearpage

\clearpage

\clearpage

\clearpage

\clearpage

\clearpage

\end{document}